\documentclass[journal]{IEEEtran}
\usepackage{amsmath,amssymb}
\usepackage{cite}
\ifCLASSINFOpdf
   \usepackage[pdftex]{graphicx}
\else
   \usepackage[dvips]{graphicx}
\fi
\usepackage{subfigure}

\usepackage{xcolor}
\usepackage{colortbl,booktabs}

\begin{document}
\title{ICINet: ICI-Aware Neural Network Based Channel Estimation for Rapidly Time-Varying OFDM Systems}
\author{
%Author 1, Author 2, and Author 3
Yi~Sun,~\IEEEmembership{Student Member,~IEEE,} Hong~Shen,~\IEEEmembership{Member,~IEEE,} Zhenguo Du, Lan Peng, and~Chunming~Zhao,~\IEEEmembership{Member,~IEEE}
% <-this % stops a space
%\thanks{Author 1, Author 2, and Author 3 are with xxx (e-mail:\{Author 1, Author 2, Author 3\}@xxx).}

\thanks{Y. Sun, H. Shen, and C. Zhao are with the National Mobile Communications Research Laboratory, Southeast University, Nanjing 210096, China (e-mail:\{sun\_yi, shhseu, cmzhao\}@seu.edu.cn). C. Zhao is also with Purple Mountain
Laboratories, Nanjing 211111, China.}

% <-this % stops a space
\thanks{Z. Du and L. Peng are with Huawei Device Co., Ltd, Shenzhen 518129, China (e-mail:\{zhenguo.du, penglan1\}@huawei.com).}

}

\maketitle

\begin{abstract}
A novel intercarrier interference (ICI)-aware orthogonal frequency division multiplexing (OFDM) channel estimation network ICINet is presented for rapidly time-varying channels. ICINet consists of two components: a preprocessing deep neural subnetwork (PreDNN) and a cascaded residual learning-based neural subnetwork (CasResNet). By fully taking into account the impact of ICI, the proposed PreDNN first refines the initial channel estimates in a subcarrier-wise fashion. In addition, the CasResNet is designed to further enhance the estimation accuracy. The proposed cascaded network is compatible with any pilot patterns and robust against mismatched system configurations. Simulation results verify the superiority of ICINet over existing networks in terms of better performance and much less complexity.
\end{abstract}


\section{Introduction}
Fifth Generation (5G) and beyond 5G (B5G) wireless communication systems are envisioned to support reliable data transmissions even under the challenging high mobility scenario \cite{7894280}. When the time-domain channel varies rapidly, the performance of the widely used orthogonal frequency division multiplexing (OFDM) modulation can be severely degraded due to the non-negligible intercarrier interference (ICI) caused by the Doppler shift. To address this problem, it is necessary to develop advanced OFDM channel  estimation schemes that can adapt to the fast time-varying channel.
%Least square (LS) and linear minimum mean square error (LMMSE) estimation, two of the conventional channel estimation methods, suffer from poor performance and high computational complexity, respectively \cite{5456443}. Moreover, these methods are based on the assumption that the channel impulse response (CIR) remains constant within the duration of one OFDM symbol. Therefore, they can hardly adapt to rapid channel variations.

Recently, deep learning has drawn widespread attentions in the area of wireless physical-layer techniques \cite{8054694,8663966,8715338}. In particular, the deep learning based OFDM channel estimation has been investigated in a number of prior works such as \cite{8052521,8509622,8761312,8933411,8640815,8944280}. As a first attempt, \cite{8052521} introduced a fully-connected neural network to recover the transmitted symbols directly, whereas the channel was estimated in an implicit way. Alternatively, concerning the channel estimation and signal detection for OFDM systems, the authors of \cite{8509622} adopted a model-driven approach by incorporating the expert knowledge into the neural network design. However, these two works did not consider time-varying channels. In \cite{8761312}, a channel estimation network called ChanEstNet was developed for high mobility scenarios using block-type pilot patterns. Furthermore, the work was extended to the multiple-input-multiple-output (MIMO) setup in \cite{8933411}. The similarity between the pilot-based channel estimation and the super resolution (SR) technique in image processing was exploited in \cite{8640815,8944280} to yield a high-quality channel estimate for OFDM systems. Treating the time-frequency response of a doubly selective channel as a two-dimensional (2D) image, the authors of [9] and [10] proposed ChannelNet based on the convolutional neural network and ReEsNet based on the residual network, respectively. It is worthwhile noting that the aforementioned neural network based channel estimation methods do not explicitly take the impact of ICI into account, which can lead to performance degradation especially in presence of rapid channel variations.

In this work, we propose a novel ICI-aware channel estimation network ICINet, which is superior to existing networks in terms of both performance and complexity. Specifically, ICINet consists of a preprocessing deep neural subnetwork (PreDNN) and a cascaded residual learning-based neural subnetwork (CasResNet). The PreDNN, which includes the ICI information on a few adjacent subcarriers into the network input, is used to refine the least square (LS) based initial channel estimates, while CasResNet can further improve the accuracy of the channel estimates over the 2D time-frequency grid. Owing to the model-driven strategy and the parameter-sharing mechanism, the proposed cascaded network is adaptive to different pilot patterns and system configurations. In addition, as a preprocessing network, the proposed PreDNN can be readily cascaded with other networks for performance enhancement with only a slight increase in the complexity.

\section{System Model}
Consider an OFDM system over doubly selective fading channels, where each subframe is constituted by $K$ subcarriers and $T$ OFDM symbols. In order to suppress the inter-symbol interference (ISI) caused by the multipath effect, a cyclic prefix (CP) with length $N_{CP} \geq N_L-1$ is added before each OFDM symbol, where $N_L$ denotes the number of channel taps. It was usually assumed that the CIR remains constant within the duration of one OFDM symbol in prior works such as \cite{8052521,8509622}, which applies for slowly time-varying channels. However, for the high mobility scenario, this assumption no longer holds \cite{1275673}. In fact, the fast channel variation destroys the orthogonality among subcarriers and leads to severe ICI. Consequently, after removing the CP and performing the discrete Fourier transform (DFT), the frequency-domain received signal of the $t$-th OFDM symbol ${{\bf{Y}}_t} \in \mathbb{C}{^{K \times 1}}$ can be expressed as
\begin{equation} 
	{{\bf{Y}}_t} = {{\bf{H}}^{(t)}}{{\bf{X}}_t} + {{\bf{W}}_t}, \quad t=1,\cdots,T,
	\label{eq1}
\end{equation}
where ${{\bf{X}}_t} \in \mathbb{C}{^{K \times 1}}$ and ${{\bf{W}}_t} \in \mathbb{C}{^{K \times 1}}$ are the frequency-domain transmitted symbols and zero-mean additive white Gaussian noise with covariance $\sigma^2 \mathbf I$, respectively. ${{\bf{H}}^{(t)}} \in \mathbb{C}{^{K \times K}}$ denotes the channel frequency response (CFR) matrix corresponding to the $t$-th OFDM symbol, which can be represented as
\begin{equation} 
	{{\bf{H}}^{(t)}} = {\bf{F}}{{\bf{G}}^{(t)}}{{\bf{F}}^H}, \quad t=1,\cdots,T,
	\label{eq7}
\end{equation}
where ${{\bf{F}}} \in \mathbb{C}{^{K \times K}}$ is the $K$-point DFT matrix and ${{\bf{G}}^{(t)}} \in \mathbb{C}{^{K \times K}}$ is the channel impulse response (CIR) matrix of the $t$-th OFDM symbol and given by
\begin{equation} 
	\begin{split}
{{\bf{G}}^{(t)}} = \left[ {\begin{array}{*{20}{c}}
		{g_{1,1}^{(t)}}&0& \cdots &{g_{1,{N_L}}^{(t)}}& \cdots &{g_{1,2}^{(t)}}\\
		{g_{2,2}^{(t)}}&{g_{2,1}^{(t)}}&0& \cdots & \cdots &{g_{2,3}^{(t)}}\\
		\vdots & \ddots & \ddots & \ddots & \ddots & \vdots \\
		0& \cdots &0&{g_{K,{N_L}}^{(t)}}& \cdots &{g_{K,1}^{(t)}}
\end{array}} \right],\\  \quad t=1,\cdots,T,
	\label{eq8}
\end{split}
\end{equation}
where ${g_{i,j}^{(t)}}$ denotes the CIR of the $j$-th channel tap at time instant $i$. 

\begin{figure}[t]
	\centering
	\includegraphics[width=5.3cm]{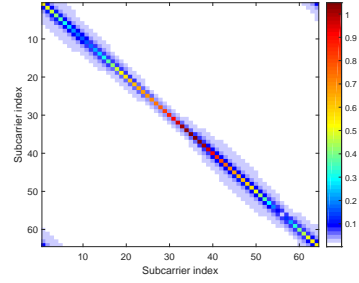}
	\caption{Illustration of the normalized magnitude of a CFR matrix ($K=64, {f_d}=0.1$) .}
	\label{fig1}
\end{figure}

Note that the diagonal elements of  ${{\bf{H}}^{(t)}}$ are the desired frequency-domain channel responses, while the non-diagonal elements all represent ICI coefficients. Accordingly, we rewrite (\ref{eq1}) into an element-wise form as 
\begin{equation} 
\begin{split}
     {Y_{k,t}} = {H_{k,k}^{(t)}}{X_{k,t}} + \sum\limits_{m = 0,m \ne k}^{K - 1} {H_{k,m}^{(t)}{X_{m,t}} + {W_{k,t}}}, \\k=1,\cdots,K,\ t=1,\cdots,T,
\label{eq2}
\end{split}
\end{equation}
where $Y_{k,t}$ is the $k$-th entry of ${{\bf{Y}}_t}$,  $X_{k,t}$ is the $k$-th entry of ${{\bf{X}}_t}$, $W_{k,t}$ is the $k$-th entry of ${{\bf{W}}_t}$, and ${H_{k,m}^{(t)}}$ is the ($k,m$)-th entry of ${\bf{H}}^{(t)}$. %refers to the ICI coefficient of the $m$-th subcarrier to the $k$-th subcarrier. 
The three terms of (\ref{eq2}) represent the useful signal, the ICI, and the noise at the $k$-th subcarrier, respectively. An example of the CFR matrix ${{\bf{H}}^{(t)}}$ is shown in Fig. \ref{fig1}, where $f_d$ denotes the maximum normalized Doppler shift. It can be seen from Fig. 1 that the ICI power is mainly concentrated within a few adjacent subcarriers. In addition, there are also some non-ignorable entries in the bottom-left corner and the top-right corner, which is due to the cyclicity of the CP.
%manifests the “cyclicity” brought by the CP.

In this work, we adopt an efficient grid pilot pattern for channel estimation, i.e., the pilots occupy $K_p$ evenly spaced subcarriers of $T_p$ nonconsecutive OFDM symbols in each subframe. Note that this kind of pilot pattern has been commonly used in OFDM based wireless communication systems.% such as 4G and 5G. 

\section{Proposed Channel Estimation Network}
\begin{figure*}[t]
	\centering
	\includegraphics[width=11.4cm]{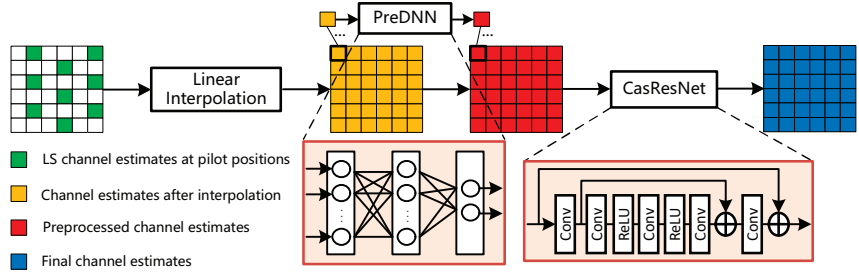}
	\caption{Block diagram of the proposed channel estimation method.}
	\label{fig2}
\end{figure*}

In this section, we develop a novel neural network based channel estimation method for the above introduced OFDM system. As depicted in Fig. \ref{fig2}, the proposed channel estimation method consists of three stages. For the first stage, we obtain initial channel estimates by performing the conventional LS estimation at the pilot symbols and linearly interpolating channels at the data symbols. For the second stage, we apply a preprocessing subnetwork PreDNN to incorporate the effect of ICI caused by the rapid channel variation. Finally, a residual learning based subnetwork CasResNet is cascaded with PreDNN to further improve the accuracy of the channel estimates. The details of the proposed two subnetworks and the corresponding training strategies are provided in the following.

\subsection{PreDNN}
It has been shown in (\ref{eq2}) that the received signal at each subcarrier is subject to ICI. However, to the best of our knowledge, the effect of ICI is not considered in existing deep learning based fast time-varying channel estimation schemes, which can lead to performance degradation. Inspired by this fact, we propose a novel subnetwork PreDNN, which takes into account the non-negligible ICI and thus can remarkably enhance the channel estimation performance.

%Instead of handling the whole time-frequency girds at a time, the application object of PreDNN is each single subcarrier, which is so-called “element-wise”. Owing to the parameter-sharing mechanism, which means that each subcarrier shares the same set of tunable parameters, the scale of the network can be efficiently reduced and independent from the size of channel response matrix.

Specifically, PreDNN is a fully-connected neural network that involves only one hidden layer, which contains 32 neurons and adopts the rectified linear unit (ReLU) as the activation function. %, i.e.\ $f(x) = \max (0,x)$.
In order to incorporate the ICI into the network design, we embrace the information of both the subcarrier of interest and its adjacent subcarriers into the input of PreDNN. Since the ICI power is significant within only a few adjacent subcarriers, we can greatly simplify the input of PreDNN. Concretely, we first perform single-tap equalization for each subcarrier and obtain the estimate of $X_{k,t}$ as
\begin{equation} 
	{{\hat X} _{k,t}} = \mathop {\arg \min }\limits_{X \in \cal{A}} {\left| {\frac{{{Y_{k,t}}}}{{{{\hat {{H}}}_{k,t}}}} - X} \right|^2}, 
	\label{eq3}
\end{equation}
where $\cal{A}$ is the modulation alphabet and $\hat{{H}}_{k,t}$ is the initial channel estimate of ${H_{k,k}^{(t)}}$ obtained in the first stage. For the $k$-th subcarrier of the $t$-th OFDM symbol, let us assume that the ICI mainly comes from $2N_{ICI}$ neighbouring subcarriers, i.e., subcarrier $k-N_{ICI}$ to $k-1$ and subcarrier $k+1$ to $k+N_{ICI}$. Then, the received signals of all the $2N_{ICI}+1$ subcarriers ${Y_{k - {N_{ICI}},t}}$, $\cdots$,${Y_{k,t}}$, $\cdots$,${Y_{k + {N_{ICI}},t}}$, the corresponding pre-estimated data symbols ${{\hat X}_{k - {N_{ICI}},t}}$, $\cdots$, ${{\hat X}_{k,t}}$, $\cdots$,${{\hat X}_{k + {N_{ICI}},t}}$, and the initial channel estimate of the current subcarrier ${{\hat H}_{k,t}}$, are merged into a column vector as the input of PreDNN. The cyclicity in the frequency domain as shown in Fig. \ref{fig1} is exploited for the first and last few subcarriers. For example, when $K=64$ and $N_{ICI} = 1$, the input for the first subcarrier would be $[Y_{64,t},Y_{1,t},Y_{2,t},{\hat X}_{64,t},{\hat X}_{1,t},{\hat X}_{2,t},{\hat {{H}}}_{1,t}]^T$. % while the counterpart for the last subcarrier would be $[Y_{63,t},Y_{64,t},Y_{1,t},{\hat X}_{63,t},{\hat X}_{64,t},{\hat X}_{1,t},{\hat H}_{64,t}]^T$.
Note that each complex element should be divided into real and imaginary parts to facilitate the use of software libraries such as Tensorflow. Hence, the input layer contains $8N_{ICI}+6$ neurons in total. Likewise, the output layer contains 2 neurons, corresponding to the real and imaginary parts of the improved channel estimate ${{\tilde {{H}}}_{k,t}}$.

We note that the proposed PreDNN is used for improving the channel estimate achieved in the first stage in a subcarrier-wise manner. In other words, once the network parameters are trained, they are shared by all subcarriers. Therefore, the scale of the network can be efficiently reduced and independent of the subframe size. Clearly, this can also be achieved by using a $1 \times 1$ convolutional layer.

\subsection{CasResNet}
After the preprocessing by PreDNN, the improved channel estimates of all the subcarriers are concatenated accordingly to form the 2D time-frequency channel response matrix $\tilde{\bf{{{H}}}} \in \mathbb{C}{^{K \times T}}$. Regarding $\tilde{\bf{{{H}}}}$ as a noisy low-resolution image (equivalently transformed into $\mathbb{R}{^{K \times T \times 2}}$), we introduce CasResNet to enhance the resolution and alleviate the effect of noise. 

CasResNet is a small-scale neural network based on the residual learning, which was first studied in \cite{7780459} to address the vanishing gradient problem and improve the learning performance. The first layer is a convolutional layer using 8 filters of size $5 \times 5 \times 2$, which is expected to extract some low-level features from the original image $\tilde{\bf {{H}}}$, while more layers are stacked subsequently to further capture high-level features. Each of the following 3 convolutional layers has 8 filters of size $3 \times 3 \times 8$, with one ReLU layer inserted between each two of them. To be consistent with the input $\tilde{\bf {{H}}}$, the last convolutional layer uses 2 filters of size $5 \times 5 \times 8$ to generate a result of size $K \times T \times 2$.  Specially, a double-residual-mapping nested architecture is built with two shortcut connections, by adding up the outputs of two certain layers. In the end, the output of CasResNet is reorganized according to the real and imaginary parts to get the final channel estimates $\breve{\bf {{H}}} \in \mathbb{C}{^{K \times T}}$.

Compared to the most relevant work ReEsNet\cite{8944280}, the superiority of the proposed CasResNet mainly lies in the following aspects:
\begin{enumerate} 
\item
The transposed convolution layer in ReEsNet for upsampling should be redesigned for a mismatched pilot pattern (including the number of pilots) or subframe size. Different from ReEsNet, we adopt the linear interpolation to implement pre-upsampling, which exploits the knowledge of traditional OFDM communications. Hence, CasResNet is model-driven and thus unaffected by the pilot pattern. 
\item
The parameter sharing property of PreDNN and the convolutional layers in CasResNet also enable our network to adapt to various subframe sizes without the need of redesigning, which guarantees the good compatibility. 
\item
By fully taking into account the impact of ICI, the preprocessing subnetwork PreDNN provides refined channel estimates for the cascaded subnetwork CasResNet. In this way, the network size of CasResNet can be reduced without compromising much performance, which thus efficiently lowers the complexity. %In this way, the extra information of ICI has been embraced into the input of CasResNet, which may help the network avoid falling into some local optimum. Obviously, the performance of the network highly depends on its input and architecture. Hence, benefiting from the good preprocessed input, the architecture of CasResNet can be relatively simple, which results in much less complexity.
\end{enumerate}

\subsection{Training Strategy}
Denote the function of PreDNN and CasResNet as ${\cal{F}_{{\rm{Pre}}}}( \cdot )$ and ${\cal{F}_{{\rm{Cas}}}}( \cdot )$, respectively. Then, the output of the proposed network can be expressed as
\begin{equation} 
{\breve {\bf{H}}} = {\cal{F}_{{\rm{Cas}}}}({\cal{F}_{{\rm{Pre}}}}({\bf{Y}},{\bf{\hat X}},{\bf{\hat {{H}}}},{{\bf{\Theta }}_{{\rm{Pre}}}}),{{\bf{\Theta }}_{{\rm{Cas}}}}),
\label{eq4}
\end{equation}
where ${\bf{Y}}$, ${\bf{\hat X}}$ and ${\bf{\hat {{H}}}}$ denotes the received signals, the pre-estimated symbols and the initial channel estimates over the 2D time-frequency channel grid. In addition, ${{\bf{\Theta }}_{{\rm{Pre}}}}$ and ${{\bf{\Theta }}_{{\rm{Cas}}}}$ are the trainable parameters of PreDNN and CasResNet, respectively. %Considering that the end-to-end training requires a large training dataset and suffers from a poor convergence rate, 
{Considering that the subnetworks PreDNN and CasResNet are designed with distinct goals as introduced in Section~III, we train them in a sequential manner instead of an end-to-end manner.} we train the two subnetworks in a sequential manner. First, we train the parameters ${{\bf{\Theta }}_{{\rm{Pre}}}}$ of PreDNN by minimizing the mean square error (MSE) between the estimated and true channel responses. The corresponding loss function is given by
\begin{equation} 
L({{\bf{\Theta }}_{{\rm{Pre}}}}) = \frac{1}{{\left| {\cal{S}} \right|}}\sum\limits_{{\bf{\bar H}} \in {\cal{S}}} {\left\| {{\tilde{\bf{{{H}}}}} - {\bf{{\bar{H}}}}} \right\|_F^2},
\label{eq5}
\end{equation}
where ${\cal{S}}$ is the set of training samples, ${\bf{{\tilde{H}}}}$ and ${\bf{{\bar{H}}}}$ are the channel estimates refined by PreDNN and the true channel responses, respectively, and $\|\cdot\|_F^2$ denotes the Frobenius norm of a matrix. Then, with ${{\bf{\Theta }}_{{\rm{Pre}}}}$ fixed, we train CasResNet using the loss function defined as
\begin{equation} 
L({{\bf{\Theta }}_{{\rm{Cas}}}}) = \frac{1}{{\left| {\cal{S}} \right|}}\sum\limits_{{\bf{\bar H}} \in {\cal{S}}} {\left\| {{\breve{\bf{{{H}}}}} - {\bf{{\bar{H}}}}} \right\|_F^2}, 
\label{eq6}
\end{equation}
where ${\bf{{\breve{H}}}}$ is the final channel estimates. Both subnetworks are trained by the well-established Adam optimizer, where the batch size, the initial learning rate, and the number of epochs are set to 200, 0.001, and 100, respectively.

%\begin{figure*}[t]
%	\subfigure[84 pilots ($K_{p}=21, T_{p}=4$)]
%	{\begin{minipage}[t]{0.33\textwidth}
%			\centering
%			\includegraphics[width=6.2cm]{PreDNN.eps}
%			\label{fig3}
%		\end{minipage}		
%	}
%	\subfigure[84 pilots ($K_{p}=21, T_{p}=4$), $N_{ICI}=2$]
%	{\begin{minipage}[t]{0.33\textwidth}
%			\centering
%			\includegraphics[width=6.2cm]{MSE_v3.eps}
%			\label{fig4}
%		\end{minipage}		
%	}
%	\subfigure[48 pilots ($K_{p}=16, T_{p}=3$), $N_{ICI}=2$]
%	{\begin{minipage}[t]{0.33\textwidth}
%			\centering
%			\includegraphics[width=6.2cm]{MSE_16_3.eps}
%			\label{fig5}
%		\end{minipage}	
%	}
%	\caption{MSE performances of different channel estimation methods.}	
%	%	\caption{{(a)MSE performances of PreDNN in terms of different $N_{ICI}$; (b) MSE performances with 84 pilots ($N_{pf}=21,N_{ps}=4$) of different channel estimation methods; (c) MSE performances with 48 pilots ($N_{pf}=16,N_{ps}=3$) of different channel estimation methods.}}
%	\label{fig}
%\end{figure*}

\begin{table}[t]
	\centering
	\caption{System Configuration}
	\renewcommand\arraystretch{1.2}
	\begin{tabular}{ccc}
		\toprule %[1pt]
		\textbf{Parameters}  & \textbf{Values}  \\
		\hline
		Carrier frequency & 2 GHz   \\
		\hline
		Subcarrier spacing &15 KHz   \\
		\hline
		Number of subcarriers
		  &128	  \\
		\hline
		Number of OFDM symbols per subframe
		 &14  \\
		\bottomrule %[1pt]
	\end{tabular}
	\label{tab1}
\end{table}
\begin{figure}[t]
	\centering
	\includegraphics[width=6.6cm]{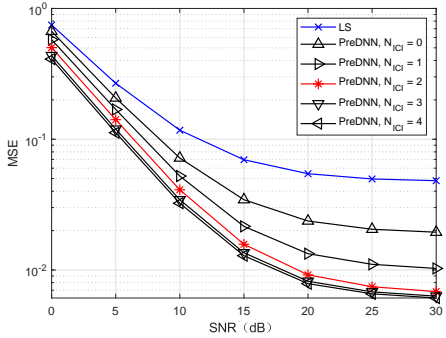}
	\caption{MSE performance of PreDNN in terms of different $N_{ICI}$'s with 84 pilots ($K_{p}=21,T_{p}=4$).}
	\label{fig3}
\end{figure}
\section{Simulation Results}
We consider an OFDM system configured as in Table \ref{tab1}. %follows: the carrier frequency and the subcarrier spacing are 2 GHz and 15 kHz, respectively, and there are $K = 128$ subcarriers and $T = 14$ OFDM symbols in a subframe. 
Quadrature phase-shift keying (QPSK) is used for modulation. For the training dataset, 12000 subframes are generated under the SNR of 10 dB with 10000 subframes for training and 2000 subframes for validation, each of which corresponds to a specific channel realization. {The linear attenuation (LA) delay profile \cite{7439864} with the Jakes Doppler spectrum is employed as the training channel model.} In order to guarantee the generalization of the proposed network, the numbers of channel taps and the maximum Doppler shifts are randomly selected from 3 to 9 and 800 to 1200 Hz, respectively. On the other hand, to demonstrate the robustness, the performance of the trained network is evaluated with 2000 testing subframes under a wide range of SNRs and a mismatched scenario, {where the Extended Vehicle A (EVA) delay profile with 6 paths and the Jakes Doppler spectrum is adopted, and the maximum Doppler shift is set to 926 Hz, i.e., a vehicle speed of 500 km/h. }

\begin{figure}[t]
	\centering
	\includegraphics[width=6.6cm]{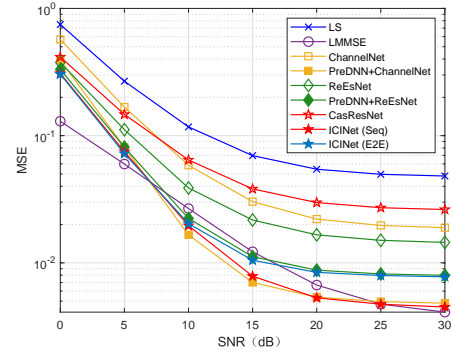}
	\caption{MSE performance of different channel estimation methods with 84 pilots ($K_{p}=21,T_{p}=4$) and $N_{ICI}=2$.}
	\label{fig4}
\end{figure}

\begin{figure}[t]
	\centering
	\includegraphics[width=6.6cm]{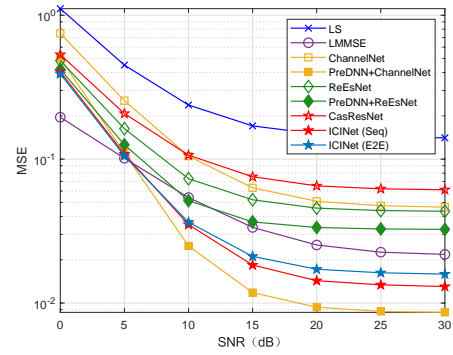}
	\caption{MSE performance of different channel estimation methods with 48 pilots ($K_{p}=16,T_{p}=3$) and $N_{ICI}=2$.}
	\label{fig5}
\end{figure}

\subsection{Performance Evaluation}
The refinement on the channel estimation by PreDNN can be partially attributed to the side information of the adjacent subcarriers. Hence, we first need to determine the number of adjacent subcarriers that should be taken into account. Accordingly, we present the MSE performance of PreDNN versus different $N_{ICI}$'s in Fig. \ref{fig3}. It can be observed that a noticeable performance gain can be achieved as $N_{ICI}$ increases from 0 to 2. After that, the performance of PreDNN gradually converges. Considering that a larger $N_{ICI}$ corresponds to a larger network and thus leads to higher complexity, we set $N_{ICI}=2$ during the simulation.%\footnote{{\color{blue}In the real unknown environment, the vehicle speed is an easily accessible indicator, based on which we can generate reasonable simulation data with the Jakes model. In particular, the higher the vehicle speed is, the more severe the ICI is, and in turn the larger $N_{ICI}$ should be set. Therefore, we can utilize the simulation data to make a look-up table of the appropriate value of $N_{ICI}$ with the corresponding speed interval in advance, which can provide a guidance in practice. }}
\footnote{{The value of $N_{ICI}$ mainly depends on the maximum Doppler shift, or, equivalently, the vehicle speed with given carrier frequency. In general, the higher the vehicle speed is, the more severe the ICI is, and the larger $N_{ICI}$ should be set. In practice, we can divide the range of all supported vehicle speeds into a few intervals and then establish a look-up table which maps the speed interval to the value of $N_{ICI}$ via offline simulations. Then, during the online channel estimation, we can first estimate the vehicle speed and then determine the value of $N_{ICI}$ using this table.}} It can also be found that even when $N_{ICI}=0$, PreDNN can still outperform the LS estimation, which indicates the powerful capability of deep neural networks.

The MSE performance of different channel estimation methods is compared in Figs. \ref{fig4} and \ref{fig5}. For the case of 84 pilots, the LS estimation suffers from a high error floor due to the rapid channel variation, whereas the deep methods can achieve much better MSE performance. In particular, our proposed network ICINet, i.e.,\ `PreDNN+CasResNet', yields a much lower MSE than ChannelNet\cite{8640815} and ReEsNet\cite{8944280}, which is also comparable to the linear minimum mean square error (LMMSE) estimation that requires the prior knowledge of channel statistics and noise variance.\footnote{{The performance of the optimal MMSE estimate ${\mathbb{E}}({\bf{\bar H}}\left| {\bf{Y}} \right.)$ is not provided considering the difficulty in obtaining an analytical form of the complicated posterior probability density function (PDF) $p({\bf{\bar H}}\left| {\bf{Y}} \right.)$ of the general doubly selective OFDM channel. Moreover, it is also intractable to calculate the MMSE estimate numerically due to the large size of the OFDM channel.}} {On the other hand, it can be seen that only using CasResNet leads to poor performance due to the limited network size, which validates the benefits from the preprocessing subnetwork PreDNN.} Furthermore, we test the combination of PreDNN and some existing networks, i.e., \ `PreDNN+ChannelNet' and `PreDNN+ReEsNet', both of which exhibit better performance than the original networks. However, the additional gain obtained by cascading PreDNN with ReEsNet is less obvious, since PreDNN can only be applied at the pilots, resulting in limited improvement. When the number of pilots is reduced to 48, the gaps between the deep learning based methods and the LS estimation become larger, and the proposed ICINet still performs considerably better than the existing networks and can even outperform the LMMSE estimation. Note that, for `PreDNN+ChannelNet', the introduction of PreDNN makes it more attractive in terms of performance, but its complexity is also much higher as presented in the following subsection.   

\begin{figure}[t]
	\centering
	\includegraphics[width=6.6cm]{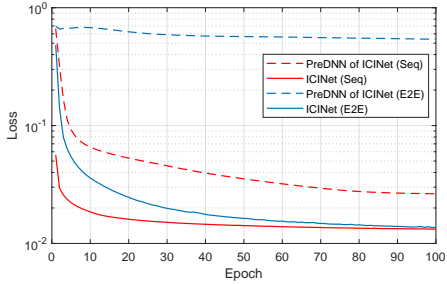}
	\caption{{Training loss comparison between the two training strategies with 84 pilots ($K_{p}=21,T_{p}=4$) and $N_{ICI}=2$.}}
	\label{fig6}
\end{figure}

{In addition, we can find that the sequentially trained ICINet, denoted by `ICINet (Seq)', shows a better generalization than the end-to-end counterpart `ICINet (E2E)' in the high SNR region, which indicates that the end-to-end training suffers overfitting due to the large number of network parameters to be optimized simultaneously. For further illustration, we compare the training loss of different training strategies in Fig.~6. Although the end-to-end training can reach almost the same loss as the sequential training, it suffers a slower convergence rate. This is due to the fact that, in the sequential training procedure, we have specified the training goal for the first subnetwork PreDNN, which can provide a good initial value for the cascaded subnetwork CasResNet and therefore facilitate its training. On the other hand, the end-to-end training regards the whole network as a black box whose training performance highly relies on the training data. In fact, it can be observed that the PreDNN under the end-to-end training does not even help improve the accuracy of the initial channel estimate.}

\subsection{Complexity Analysis}

\begin{table}[t]
	\centering
	\caption{Complexity of Different Channel Estimation Networks}
	\renewcommand\arraystretch{1.2}
	\begin{tabular}{ccc}
		\toprule %[1pt]
		 \textbf{Network}  & \textbf{MAC Operations} & \textbf{Network Parameters} \\
		\hline
		ChannelNet & 2.41e9 &676354  \\
		\hline
		PreDNN+ChannelNet &2.41e9   &677156\\
		\hline
		ReEsNet  &8.15e6	  &44674\\
		\hline
		PreDNN+ReEsNet &9.53e6  &45476 \\
		\hline
		CasResNet &4.53e6  &2562 \\
		\hline
		\textbf{ICINet} &\textbf{5.91e6} &\textbf{3364}  \\ 
		\bottomrule %[1pt]
	\end{tabular}
    \label{tab2}
\end{table}

%The complexity of LS estimation and the aforementioned networks are all ${\cal{O}}(KT)$, while the complexity of LMMSE estimation is ${\cal{O}}(KT({K_p}{ T_p })^2+({K_p}{ T_p })^3)$, which grows cubically with the size of a subframe \cite{8944280}. More quantitatively, t
The complexity of different networks is compared in terms of the numbers of multiply-accumulate (MAC) operations and network parameters, which characterize the computational complexity and the memory usage, respectively. The results listed in Table \ref{tab2} show that ICINet not only requires the fewest MAC operations but also has the fewest network parameters, with advantages of up to several orders of magnitude over ChannelNet. Besides, it can be found that the combination with PreDNN  does not bring much burden to the original networks.  

\section{Conclusion}
In this letter, we have developed a novel OFDM channel estimation network for rapidly time-varying channels by incorporating  the ICI into the network design. The proposed network ICINet consists of a preprocessing subnetwork PreDNN cascaded with a residual learning based subnetwork CasResNet, which can remarkably improve the accuracy of existing neural network based channel estimation schemes while with reduced computational complexity and memory usage. Moreover, the proposed network also exhibits good compatibility and robustness. % of the proposed network makes it more competitive to other methods. 

%\appendices

%\section{}

%\section*{Acknowledgment}

\bibliography{myBibFile}

\end{document}